\documentclass[conference]{IEEEtran}
\IEEEoverridecommandlockouts
\usepackage{cite}
\usepackage{amsmath,amssymb,amsfonts}
\usepackage{graphicx}
\usepackage{textcomp}
\usepackage{xcolor}
\def\BibTeX{{\rm B\kern-.05em{\sc i\kern-.025em b}\kern-.08em
    T\kern-.1667em\lower.7ex\hbox{E}\kern-.125emX}}

\usepackage[T1]{fontenc}
\usepackage{balance}

\definecolor{darkblue}{HTML}{09357A}
\usepackage{hyperref}

\usepackage{algorithm}
\usepackage{algorithmicx}
\usepackage{algpseudocode}

\usepackage[autostyle=false, style=english]{csquotes}
\MakeOuterQuote{"}

\ifCLASSOPTIONcompsoc
\usepackage[caption=false,font=normalsize,labelfont=sf,textfont=sf]{subfig}
\else
\usepackage[caption=false,font=footnotesize]{subfig}
\fi

\usepackage[most]{tcolorbox}  %

\newcommand{\altparagraph}[1]{%
  \smallskip
  \textbf{#1}
  \noindent
}

\newtcolorbox{definitionbox}{
    enhanced,                           %
    colback=gray!10,                    %
    colframe=gray!10,                   %
    left=5mm,                           %
    right=5mm,                          %
    borderline west={1mm}{0mm}{gray},   %
}
\newenvironment{definition}[1]{
    \begin{definitionbox}
    \textbf{Definition: #1} \\[0.5em]
}{
    \end{definitionbox}
}

\newcommand\copyrighttext{
  \footnotesize \textcopyright~2025 IEEE. Personal use of this material is permitted. Permission from IEEE must be obtained for all other uses, in any current or future media, including reprinting/republishing this material for advertising or promotional purposes, creating new collective works, for resale or redistribution to servers or lists, or reuse of any copyrighted component of this work in other works. This article is published in the 2025 IEEE 18th International Conference on Cloud Computing (CLOUD). DOI: \href{https://doi.org/10.1109/CLOUD67622.2025.00051}{\textcolor{darkblue}{10.1109/CLOUD67622.2025.00051}}.}
\newcommand\copyrightnotice{
\begin{tikzpicture}[remember picture,overlay]
\node[anchor=south,yshift=10pt] at (current page.south) {\fbox{\parbox{\dimexpr\textwidth-\fboxsep-\fboxrule\relax}{\copyrighttext}}};
\end{tikzpicture}
}

\begin{document}

\title{Universal Workers: A Vision for Eliminating Cold Starts in Serverless Computing}

\author{
    \IEEEauthorblockN{Saman Akbari\IEEEauthorrefmark{1}, Manfred Hauswirth\IEEEauthorrefmark{1}\IEEEauthorrefmark{2}}
    \IEEEauthorblockA{\IEEEauthorrefmark{1}Technische Universität Berlin, Open Distributed Systems, Berlin, Germany}
    \IEEEauthorblockA{\IEEEauthorrefmark{2}Fraunhofer Institute for Open Communication Systems (FOKUS), Berlin, Germany \\
                      Email: \texttt{\{akbari, manfred.hauswirth\}@tu-berlin.de}}
}

\maketitle
\copyrightnotice

\begin{abstract}
Serverless computing enables developers to deploy code without managing infrastructure, but suffers from cold start overhead when initializing new function instances. Existing solutions such as "keep-alive" or "pre-warming" are costly and unreliable under bursty workloads.
We propose universal workers, which are computational units capable of executing any function with minimal initialization overhead. Based on an analysis of production workload traces, our key insight is that requests in Function-as-a-Service (FaaS) platforms show a highly skewed distribution, with most requests invoking a small subset of functions. We exploit this observation to approximate universal workers through locality groups and three-tier caching (handler, install, import).
With this work, we aim to enable more efficient and scalable FaaS platforms capable of handling diverse workloads with minimal initialization overhead.
\end{abstract}


\section{Introduction}
\label{sec:introduction}

Function-as-a-Service (FaaS) is a cloud computing model where the platform manages the underlying infrastructure to execute functions, handling tasks like provisioning, auto-scaling, and scheduling.
This model offers developers a "serverless" experience that is becoming increasingly popular over unmanaged services such as Infrastructure-as-a-Service (IaaS)~\cite{hendrickson2016serverless}.
FaaS solutions are available from all major cloud providers, e.g., AWS Lambda or Azure Functions,
with a wide range of use cases from simple utility functions to complex workflows.

FaaS platforms start and stop function instances based on demand. Although this elasticity is a benefit of serverless computing, starting new instances introduces the cold start problem: Each launch requires loading the function, satisfying its dependencies, and setting up an execution environment, such as a virtual machine or container.

High-level languages such as Python and JavaScript account for 84\% of serverless applications~\cite{alibaba2021}, which simplifies development but comes at the cost of much higher initialization overhead compared to low-level languages like C or Rust.
As applications grow in complexity, developers also increasingly rely on third-party libraries and shared code to speed up development, which the platform must load, install, and import at startup.
This initialization overhead can significantly degrade performance.
For providers, the time spent on initialization is also wasted, because they only bill customers for execution time.

To reduce cold starts, a common approach is "keep-alive," where platforms keep recently used instances idle for several minutes or hours after execution to be reused for subsequent requests~\cite{wang2018peeking}.
Another technique is "pre-warming," where platforms proactively initialize function instances before they are needed~\cite{vahidinia2022mitigating}. 
However, both strategies are ineffective at handling bursty workloads, where sudden spikes in demand, which are not uncommon, still trigger cold starts. 
Keeping instances idle is much more expensive than creating new instances and increases infrastructure costs.
Furthermore, pre-warming only masks the underlying problem of high initialization overhead and fails to achieve high steady-state throughput~\cite{oakes2018sock}.

This paper proposes \textit{universal workers}, where computational units in a FaaS platform can execute any function with minimal initialization overhead.
We show that approximating universal workers is feasible by exploiting the highly skewed popularity of functions in FaaS platforms, where a small subset of functions receives the majority of requests.
Our work builds on previous research in XFaaS~\cite{sahraei2023xfaas}, Meta's hyperscale private cloud that optimizes hardware utilization and resource provisioning in large-scale serverless environments.

Our proposal makes the following contributions: We
examine the lifecycle of function instances to break down the
latency of cold starts, and analyze production workload traces from different FaaS platforms (Section~\ref{sec:background}). Based on this, we devise an approach to approximate universal workers using locality groups, which partition a subset of functions to a subset of workers, and three-tier caching (Section~\ref{sec:universal_workers}). We evaluate the feasibility of universal workers (Section~\ref{sec:evaluation}) and conclude with a discussion of future work (Section~\ref{sec:conclusion_future_work}).

\section{Background}  
\label{sec:background}

\subsection{Function Lifecycle}
Function instances go through three phases: initialization, invocation, and shutdown. After invocation, instances optionally enter an idle state for a period of time, where they remain available for subsequent requests to avoid the need for reinitialization. Cold starts occur when the platform needs to initialize a new function instance for invocation.

During initialization, the platform first loads the function code. Next, dependencies are resolved in three stages: First, it downloads them from package registries, e.g., \textit{PyPI} for Python packages or \textit{npm} for Node.js modules. Second, it installs these dependencies, which may involve extracting files from compressed formats or compiling native extensions. Finally, it imports the dependencies, which executes initialization code and loads the required libraries.

\begin{figure}[!th]
    \centering
    \includegraphics[width=\columnwidth]{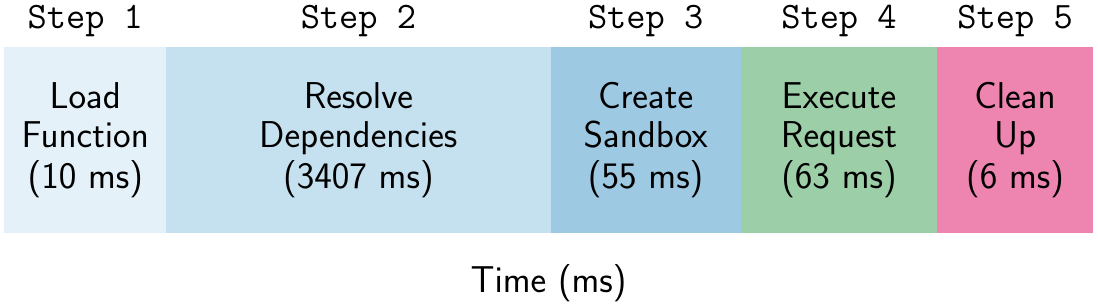}
    
    \caption{Latency breakdown of a function running the linpack benchmark (n=1000) on OpenLambda.}
    \label{fig:latency_breakdown}
\end{figure}

Cold start overheads can slow down requests significantly.
We measured the performance of the \texttt{linpack} benchmark~\cite{kim2019functionbench} ($n=1000$), which uses the NumPy package, on the open-source FaaS platform OpenLambda~\cite{hendrickson2016serverless}. Figure~\ref{fig:latency_breakdown} shows that initialization required a total of 3472 ms, whereas execution took only 63 ms and shutdown 6 ms.

\subsection{Skew in Function Popularity}
\begin{figure}[!th]
    \centering
    \includegraphics[width=\columnwidth]{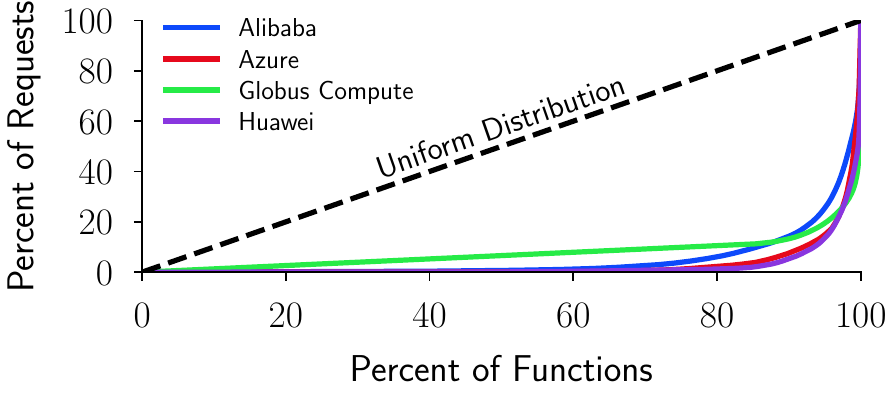}

    \caption{Skew in function popularity. A small number of functions account for the majority of requests.}
    \label{fig:skewed_function_popularity}
\end{figure}

Requests in FaaS platforms often follow a highly skewed distribution, with most requests concentrated in a small subset of functions.
In Figure~\ref{fig:skewed_function_popularity}, we analyzed publicly available workload traces from different FaaS platforms: Alibaba Cloud Function Compute~\cite{alibaba2021}, Azure Functions~\cite{zhang2021faster}, Globus Compute~\cite{bauer2024globus}, and Huawei YuanRong~\cite{joosen2025serverless}. All four platforms show a highly skewed distribution of requests.
For example, on Azure Functions, the top 0.94\% of most frequently invoked functions handle 50\% of requests and 3.54\% of functions handle 80\% of requests.

\section{Universal Workers}
\label{sec:universal_workers}
Our goal is to minimize the impact of cold starts in serverless computing. Ideally, workers should be able to execute \textit{any} function with minimal initialization overhead.
We refer to this ideal as universal workers:

\begin{definition}{Universal Worker}
    A universal worker is a computational unit in a function-as-a-service platform, capable of executing any function with minimal initialization overhead.
\end{definition}

Given that FaaS platforms run thousands of functions, it is impractical to create universal workers for every function. Instead, we exploit the skewed function popularity in FaaS platforms to approximate universal workers for popular functions that handle the majority of requests.
We propose locality groups and three-tier caching for this approximation.

\subsection{Locality Groups}
Workers in FaaS platforms have limited memory and disk capacity, which makes it infeasible to maintain a cache for all popular functions.
To address these physical constraints, we introduce locality groups:

\begin{definition}{Locality Group}
    A locality group is a subset of functions partitioned to a subset of workers.
\end{definition} 

\begin{figure}[!th]
    \centering
    \includegraphics[width=0.89\columnwidth]{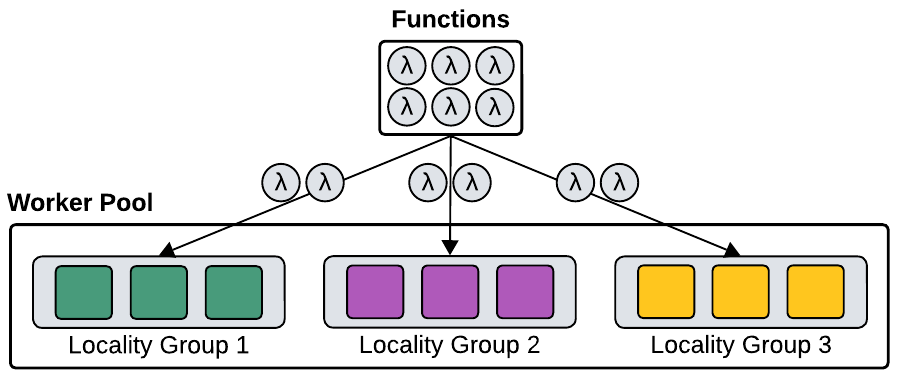}
    
    \caption{Locality group assignment. Functions are distributed across locality groups that have a pool of workers.}
    \label{fig:locality_groups}
\end{figure}

We show the formation of locality groups using an example in Figure~\ref{fig:locality_groups}. Locality groups partition both functions and workers into subsets, effectively distributing cache requirements and computational load across the platform while maintaining specialized worker pools for different function types.

\altparagraph{Partition Problem:}
Partitioning both functions and workers is a non-trivial optimization problem.
An initial implementation of locality groups could assign each runtime type to its own group to improve cache efficiency. For example, Python workers would not execute JavaScript functions. Functions can then be distributed in a round-robin fashion across groups.
For more advanced partitioning, we consider the following points important:
First, functions with shared dependencies should be grouped together.
Second, locality groups containing popular or resource-intensive functions should receive proportionally more workers.
Third, locality groups should be updated periodically to account for changes in popularity and execution patterns.

\altparagraph{Locality-Aware Routing:}
We then implement locality-aware routing to forward requests to workers within the appropriate group. Within a locality group, a scheduler component selects the specific worker to handle the request.

\subsection{Three-Tier Cache}
To eliminate the cold start overhead for functions within a locality group, we use a three-tier caching system on workers building on SOCK~\cite{oakes2018sock}.

\altparagraph{Handler Cache:} The first tier maintains idle function instances in a paused state \textit{in memory}. This is equivalent to the "keep-alive" strategy commonly found in FaaS platforms~\cite{wang2018peeking}. Unpausing is faster than creating a new instance. Paused functions do not consume CPU, but do consume memory.

\altparagraph{Install Cache:} The second tier consists of a set of pre-installed packages stored \textit{on disk}. These packages are mapped read-only into each worker's environment.

\begin{figure}[!th]
    \centering
    \includegraphics[width=0.59\columnwidth]{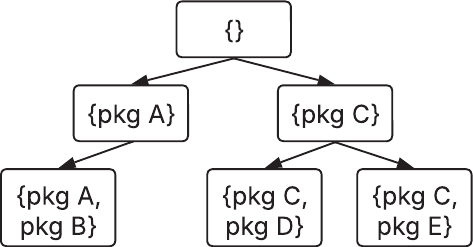}
    
    \caption{Import cache. A tree-based caching system where each node represents a sleeping process with pre-imported packages.}
    \label{fig:cache}
\end{figure}

\altparagraph{Import Cache:} The third tier implements a tree-based caching system for pre-imported packages \textit{in memory} (see Figure~\ref{fig:cache}). Each node is a sleeping process, with the root node containing only the runtime environment. Child nodes inherit pre-imported packages from parents and can import additional ones. New function instances with pre-imported packages can fork from sleeping processes within milliseconds.

\subsection{Implementation Details}
For locality groups, we implement a graph-based clustering algorithm that analyzes function dependency overlap to determine groupings.
For three-tier caching, the handler cache maintains idle function instances in a paused state using Linux's \texttt{cgroup.freeze} file; the import cache uses Linux's \texttt{fork()} with copy-on-write semantics; and the install cache uses bind mounts to share read-only packages across workers.
Integration into existing FaaS platforms requires modifications to their container systems and scheduler components.

\section{Evaluation}
\label{sec:evaluation}
We evaluate the feasibility of approximating universal workers through simulation using the four production workload traces from Section~\ref{sec:universal_workers} over a full day of request data with a total of 798,075 requests and 5,266 unique functions. For simplicity, we assume a 256 MB cache after a function's invocation to skip initialization overhead for subsequent requests. We want to answer two research questions:

\begin{itemize}
    \item Can we approximate universal workers?
    \item What are the memory requirements for the cache?
\end{itemize}

\begin{figure}[!t]
    \centering
    \includegraphics[width=0.85\columnwidth]{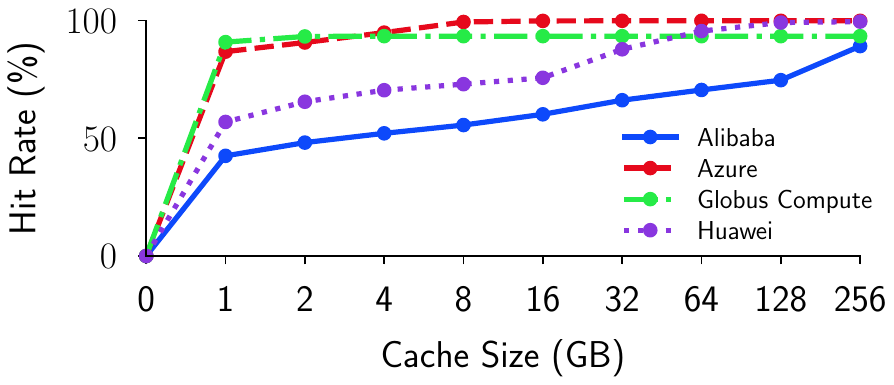}
    
    \caption{Cache hit rates. Even small cache sizes achieve high rates due to skewed function popularity, demonstrating the feasibility of universal workers.}
    \label{fig:universal_workers_feasibility}
\end{figure}

Figure~\ref{fig:universal_workers_feasibility} shows the cache hit rates by cache size using a least recently used (LRU) eviction policy. Even with a small 1 GB cache without locality groups, we achieve hit rates of 42.5\% to 90.8\% across the various FaaS platforms studied due to the skewed function popularity,
although the required cache size can be reduced by creating locality groups. Achieving close to 100\% cache hit rates on all four platforms would require a 256 GB cache, which is impractical. This again demonstrates the need for locality groups to specialize workers for a subset of functions.
Our proposed techniques of locality groups and three-tier caching could largely eliminate cold start overhead from FaaS platforms.

\section{Conclusion \& Future Work}
\label{sec:conclusion_future_work}

This paper discussed the idea of universal workers for approximate elimination of cold starts in serverless computing by exploiting the skewed function popularity in FaaS platforms. We proposed locality groups and three-tier caching to best approximate universal workers.
We hope to enable more efficient FaaS platforms with minimal overhead,
and are actively working on implementing this vision within an open-source FaaS platform. The code used in this paper is publicly available for reproducibility: \textcolor{darkblue}{\url{https://zenodo.org/records/15424821}}.

\balance
\bibliographystyle{bibliography/IEEEtran}
\bibliography{bibliography/IEEEabrv,bibliography/IEEEbibliography}

\end{document}